\documentclass[]{pasj02}

\Received{}
\Accepted{}
 
\usepackage{natbib}
\bibpunct[:]{(}{)}{,}{a}{}{,}
\usepackage{graphicx}	
\usepackage{comment}    
\usepackage{bm}
\usepackage{color}
\usepackage{CJKutf8}    
\usepackage{url}
\usepackage{caption}
\usepackage[normalem]{ulem}

\def\tc#1{\textcolor{cyan}{#1}}

\begin{document}
\begin{CJK}{UTF8}{ipxm}

\title{ 
Black Hole Spin Estimation with Time-variable Image of M87 During the Flaring State }

\author{Mikiya M. \textsc{Takahashi}\altaffilmark{1}}
\altaffiltext{1}{National Institute of Technology, Tokyo College, 1220-2, Kunugida-machi, Hachioji, Tokyo, 193-0997, Japan}
\email{m\_takahashi@tokyo-ct.ac.jp}

\author{Tomohisa \textsc{Kawashima},\altaffilmark{2}}
\altaffiltext{2}{National Institute of Technology, Ichinoseki College, Takanashi, Hagisho, Ichinoseki, Iwate, 021-8511, Japan}

\author{Ken \textsc{Ohsuga}\altaffilmark{3}}
\altaffiltext{3}{Center for Computational Sciences, University of Tsukuba, 1-1-1, Ten-nodai, Tsukuba, Ibaraki, 305-8577,
Japan}


\maketitle

\begin{abstract}
By investigating the time-variable 230 GHz images using ray-tracing general relativistic radiative transfer calculation, we propose a novel method for estimating the spin parameter of the supermassive black hole at the M87 center by utilizing the sudden and short-term increase in emissivity in the innermost region of the accretion disk. It is found that the flux of the photon ring increases simultaneously as the flux of the direct ring, which brightens first, decreases, and then gradually diminishes, when the increase in emissivity persists for $15 t_{\rm g}$ with $t_{\rm g}$ being the light crossing time of the gravitational radius. The direct ring is formed by photons emitted from the vicinity of the innermost region of the disk and traveling directly to the observer without orbiting around the black hole, while the photon ring is formed by photons passing near the spherical photon orbit. The time-averaged width of the dark region between the direct ring and the photon ring (dark crescent) becomes thinner when the spin parameter is higher and the increase in the emissivity of the accretion disk is greater. The time variation of two rings also causes the intensity-weighted center to oscillate both in the direction of the black hole's angular momentum vector projected onto the screen ($Y$-direction) and in the perpendicular direction ($X$-direction). The amplitude of oscillatory time variation in the $X$-direction becomes large when the spin parameter is higher, and that in the $Y$-direction becomes large when the increase in the emissivity of the disk is large. The spin parameter can be estimated by combining the time-averaged dark crescent width and the ratio of the amplitudes in the $X$- and $Y$-directions. This method is applicable when the duration of the increase in emissivity of the accretion disk ranges at least from approximately 10-20 $t_{\rm g}$.
\end{abstract}
\section{Introduction}
The Event Horizon Telescope (EHT) successfully observed the photon ring considered to be generated by the supermassive black hole (BH) in the center of M87 \citep{EHT1, EHT2, EHT3, EHT4, EHT5, EHT6, EHT7, EHT8}. By comparing these observations 
with results of general relativistic magnetohydrodynamics (GRMHD) simulations
, 
they inferred that the dimensionless black hole spin parameter is nonzero 
and that the angular momentum vector of the BH is oriented towards Earth
\citep{EHT1}. 
However, the constraint on the spin parameter was not successful.
In addition, although \citet{Cui2023} discovered the precession of the M87 jet which is probably caused by the frame-dragging,
they have not yet constrained the spin parameter either.

Theoretical studies regarding the determination of the spin parameter have been actively performed. In \citet{Nakamura2018}, they conducted GRMHD simulations with the spin parameter in the range from 0.5 to 0.99 and have shown 
that the position of the jet separation surface, which is 
the boundary between 
inflows and outflows, 
depends on the spin parameter. Then, general relativistic radiative transfer (GRRT) simulations by \citet{Kawashima2021, Ogihara2024} have indicated that the spin parameter could be constrained with the ring-like image generated by emissions from the separation surface. 
In \citet{Chael2021}, they constructed the semi-analytic model based on the GRMHD simulations and have suggested that the spin parameter, black hole mass, and viewing angle can be estimated using differences between the mean radius and the image centroid of the photon ring and "inner shadow" (dark region whose edge lies near the gravitational lensed event horizon smaller than the photon ring, also referred to as the "equatorial horizon" in their paper). However, this study is based 
on the assumption that the photons are emitted at 
the equatorial plane.
Methods using the image size or the multi-wavelength spectrum based on 
GRMHD simulations and semi-analytic models \citep{Drappeau2013, Chan2015, Ryan2018}
have also been suggested. \cite{Shcherbakov2012, Palumbo2020} have proposed an approach using 
the polarization of the synchrotron emission. 
Although these previous works 
focused on the synchrotron self-absorption (SSA) thin situation, it has been pointed out that the M87 core can be partially SSA-thick \citep{Kino2015}.

By calculating images considering a situation where increasing the disk emissivity (absorption coefficient)
leads to the disk becoming SSA-thick, 
\citet[][ hereafter TK19]{Kawashima2019} have suggested that the spin parameter can be estimated by observing the width of 
the crescent-shaped dark region called the "dark crescent" (hereafter DC) which is a gap 
between the photon ring formed by photons traversing paths close to the spherical photon orbit \citep{Bardeen1973, Teo2003, Takahashi2004} and the SSA-thick ISCO ring produced by photons reaching the observer without orbiting around the BH. 
However, 
it is considered that
increases in the fluxes of two rings are not 
observed simultaneously
if the duration of the disk emissivity increases is shorter than the time lag which arises from the difference in the path lengths of photons that produce the two rings.
Such a situation is entirely possible.
The difference in the photon path lengths is on the same order as the size of the spherical photon orbit, and therefore, the time lag is comparable to the dynamical time scale in the innermost region of the disk.
Since the variation time scale of the disk luminosity is also considered to be almost the same as this dynamical time scale by the observation of the radio power or the images \citep{EHT1, Philipp2022}, a study considering the time-variation of the flux of both rings is necessary. 
Spin estimation methods using 
intensity variations 
have 
been proposed by \cite{Moriyama2016, Moriyama2017, Moriyama2019, Tiede2020, Wong2021, Wong2024}.

In this work, we investigate the time-variation of simulated images using the time-dependent GRRT calculation while considering that the emissivity of the accretion disk around the supermassive BH in the center of M87 increases once 
(a part of the accretion disk becomes SSA-thick)
and then decreases. 
This leads to variations in the DC width and the intensity-weighted center. 
We propose to use these for estimating the black hole spin parameter.
In Section \ref{sec:GRRT}, we show the method of the ray-tracing GRRT calculation and the setting of the accretion disk. 
In Section \ref{sec:result}, we explain the time variation of simulated images and suggest a novel method to estimate the black hole spin parameter. 
Section \ref{sec:summary} presents summary 
and future issues. 
In this work, 
we use the GRRT code: \texttt{CARTOON} \citep{Takahashi2022} and focus on the observed frequency of 230 GHz.
We use the units $c=G=h=1$. 
Greek indices $\mu,\rho,\sigma$ denote $0-3$ and Latin indices $l,m$ denote $1-3$.

\section{Basic Equations and Numerical Setup} \label{sec:GRRT}
\subsection{Geodesic and radiative transfer equation } 
The geodesic equations in the BH space-time are denoted as follows \citep{Takahashi2017, Takahashi2022}:
\begin{align} \label{geoeq}
\begin{aligned}
    &\frac{dx^\mu}{d\lambda} = p^\mu, \\
    &\frac{d p_\mu}{d\lambda} = -\frac{1}{2} \frac{\partial g^{\rho \sigma}}{\partial x^\mu}p_\rho p_\sigma,
\end{aligned}
\end{align}
where $x^\mu$, $\ p^\mu (p_\mu)$, $\lambda$, and $g_{\rho \sigma} (g^{\rho \sigma})$ are the coordinates of the photon, the four momentum of the photon, the affine parameter, and the metric tensor, respectively.
In this paper, we use the spherical Boyer--Lindquist coordinate ($t$, $r$, $\theta$, $\phi$), whose metric is given by
\begin{align} \label{metric1}
    ds^2 &= g_{\rho \sigma} dx^\rho dx^\sigma \nonumber \\
         &= -\alpha^2 dt^2 + \gamma_{lm} (dx^l + \beta^l dt) (dx^m + \beta^m dt),
\end{align}
where $ds$ is the line element. The non-zero components of the lapse function $\alpha$, the shift vector $\beta^l$, and the spatial metric $\gamma_{lm}$ are
\begin{align} \label{metric2}
    &\alpha = \left( 1+\frac{2Mr}{\Sigma} \right)^{-\frac{1}{2}},\ \beta^r = \frac{2Mr}{\Sigma + 2Mr}, \nonumber \\
    &\gamma_{rr} = 1+\frac{2Mr}{\Sigma},\ \gamma_{\theta \theta} = \Sigma,\ \gamma_{\phi \phi} = \frac{A \sin^2 \theta}{\Sigma},\nonumber \\
    &\gamma_{r \phi} = \gamma_{\phi r} = -a\sin^2 \theta \left( 1+\frac{2Mr}{\Sigma} \right),
\end{align}
and 
\begin{align} \label{metric3}
&\Sigma = r^2+a^2\cos^2\theta, \nonumber \\
&A=\left( r^2+a^2 \right)^2-a^2\Delta \sin^2 \theta,\nonumber \\
&\Delta = r^2-2Mr+a^2,
\end{align}
where $M$ and $a$ are the BH mass and the dimensionless BH spin parameter. 

The GRRT equation is described as follows \citep{Takahashi2022}:
\begin{align} \label{grrteq}
    \frac{d\mathcal{I}}{d\lambda} = \mathcal{E} - \mathcal{AI},
\end{align}
where $\mathcal{I}, \mathcal{E}$ and $\mathcal{A}$ are the invariant intensity, invariant emissivity and invariant absorption coefficient. These invariant variables satisfy the following relations:
\begin{align} 
    &\mathcal{I} = \frac{I_\nu}{{\nu}^3}, \label{huhenI} \\
    &\mathcal{E} = \frac{j_{\nu}}{{\nu}^2}, \label{huhenE} \\
    &\mathcal{A} = \nu \kappa_{\nu} \label{huhenA},
\end{align}
where $I_{\nu}$, $j_{\nu}$ and $\kappa_{\nu}$ are the intensity, the emissivity, and the absorption coefficient measured in the local Minkowski frame at the frequency $\nu$. 
Here, $d\lambda$ is related to $ds$ as $ds = \nu d\lambda$.

In this study, we employ the angle-averaged synchrotron emissivity as
\begin{align}  
    j_{\nu} = \frac{n_{\rm e} e^2 \nu}{\sqrt{3} K_2(\Theta_{\rm e}^{-1})} \mathcal{M}(w_{\rm cs}), \label{thsync_emissivity}
\end{align}
where 
$n_{\rm e}$ is the electron number density, $e$ is the electric charge, and $K_2$ is the modified Bessel function of the second kind \citep{Mahadevan1996, Pandya2016, Kawashima2023},
dimensionless electron temperature $\Theta_{\rm e}$ is defined as $\Theta_{\rm e}\equiv k_B T_{\rm e}/m_{\rm e}$ with $k_B$ being the Boltzmann constant, $T_{\rm e}$ being the electron temperature, and $m_{\rm e}$ being the electron mass.
In addition, $\mathcal{M}(w_{\rm cs})$ is given by
\begin{align}
    \mathcal{M}(w_{\rm cs}) &= \frac{4.0505}{w_{\rm cs}^{1/6}} \left( 1 + \frac{0.4}{w_{\rm cs}^{1/4}} + \frac{0.5316}{w_{\rm cs}^{1/2}} \right) \exp (-1.8899 w_{\rm cs}^{1/3}),  
\end{align}
where $w_{\rm cs}$ is the frequency normalized 
by $\nu_{\rm cs} \left(\equiv  3eB_{\rm mag} \Theta_{\rm e}^2/4\pi m_{\rm e} \right)$
with $B_{\rm mag}$ being
the magnetic field strength measured in the fluid rest frame.
The absorption coefficient obtained by 
Kirchhoff’s law for the synchrotron emission from the thermal electron,
$\kappa_{\nu} = j_{\nu}/B_\nu$,
is used, where $B_\nu$ is the blackbody intensity. 

We use the ray-tracing GRRT code: \texttt{CARTOON} to generate geodesics by numerically integrating Equation (\ref{geoeq}) backward in time from the observer's screen.
The interval along the geodesic, $\Delta s=\nu \Delta \lambda$, is set to be $10^{-2}\  r_{\rm g}$, where $r_{\rm g}( = M)$ is the gravitational radius.
Here, the observer's viewing angle is assumed to be $30^\circ$ measured from the BH spin angular momentum axis as in TK19, and the screen is located at $r=10^3\ r_{\rm g}$.
The screen size is $-10\ r_{\rm g} \leq X,Y \leq 10\ r_{\rm g}$ which is divided into $512 \times 512$ uniform pixels and the numerical integration of the geodesic is started from the center of each pixel.
Here, $X$ and $Y$ are the horizontal and vertical axis on the screen, and 
$+Y$ direction represents the direction of the angular momentum vector of the BH projected onto the screen.
The BH mass and the distance to M87 from the Earth are adopted as $M=6.2\times 10^9 M_\odot,\ D=16.7{\rm Mpc}$ \citep{Gebhardt2009, Bird2010, Gebhardt2011}.
The observed frequency is assumed to be 230 GHz,
and the GRRT equation is integrated along the geodesics
with the interval of $10^{-2}\ r_{\rm g}$.
After the emissivity of the accretion disk increases at $t=0$ (discussed in the next subsection), the time when the first photon arrives at the screen is $t=985 t_{\rm g}$. In our simulations, we perform the GRRT calculations by considering the time lag from the accretion disk to the screen (so called slow-light calculation \citep[][]{Dexter2010, Wong2021, Murchikova2022, Wong2024}).

\subsection{Accretion disk} \label{subsec:model}
We employ a semi-analytic model \citep{Falcke2000, Broderick2006, Pu2016, Kawashima2019}, which is based on the self-similar solution of the Advection Dominated Accretion Flow \citep{Narayan1994}, and in which the electron number density ($n_{\rm e}$), electron temperature ($T_{\rm e}$), and magnetic field strength ($B_{\rm mag}$) are given by 
\begin{align}
    \label{neini}
    n_{\rm e} &= F(t) n_{\rm e}^0 \left( \frac{r}{r_{\rm g}} \right)^{-1.5} \exp \left( - \frac{r^2 \cos^2 \theta}{2H^2} \right) , \\
    \label{Teini}
    T_{\rm e} &= T_{\rm e}^0 \left( \frac{r}{r_{\rm g}} \right)^{-1},  \\
    \label{Bini}
    \frac{B_{\rm mag}^2}{8\pi} &= \frac{n_{\rm e} m_{\rm p}}{\beta_{\rm mag}} \frac{r_{\rm g}}{6r},
\end{align}
in the region $r_{\rm H} < r \leq 15\ r_{\rm g}$,
where 
$r_{\rm H}$ is the event horizon radius. 
We assume a vacuum ($j_\nu=\kappa_\nu=0$)
in the region of $r > 15\ r_{\rm g}$.
We note that, even if we do not assume a vacuum, the emissivity in the region of $r \gtrsim 15 r_{\rm g}$ is very low and does not affect the resultant simulated images.
Here, $H$, $\beta_{\rm mag}$, and $m_{\rm p}$ are the typical scale height of the disk, magnetization of the accretion disk, and proton mass. In this paper, we assume $T_{\rm e}^0 = 1.5 \times 10^{11} {\ {\rm K}}$, $H = 0.1\times r \sin \theta$, and $\beta_{\rm mag} = 0.3$ following \citet{Kawashima2019}, taking into account the structure similar to the magnetically arrested disks (e.g., \citet{Igumenshchev2003, Tchekhovskoy2011, Narayan2022}), i.e., geometrically thinner and relatively strongly magnetized disks. 
We also set $n_{\rm e}^0$ 
according to the spin parameters as follows: 
$n_{\rm e}^0= 6.5 \times 10^5 {\rm cm^{-3}}$ for $a=0.998$,
$8.0 \times 10^5{\rm cm^{-3}}$ for $a=0.75$,
and $9.0 \times 10^5{\rm cm^{-3}}$ for $a=0.5$.
With these choices of $n_{\rm e}^0, T_{\rm e}^0,$ and $\beta_{\rm mag}$, our model is broadly consistent with EHT observations. 
The initial radio flux 
in our models
is approximately 0.3 Jy 
, consistent with the measured total flux of 0.2–1.2 Jy. In the main emission region ($r\simeq1-5\ r_{\rm g}$), the electron number density is 
$6×10^4-9×10^5 {\rm \ cm}^{-3}$, the electron temperature is $3×10^{10}-1.5×10^{11}$ K.
These values lie within the observational constraints inferred from polarization measurements, which provide estimates of the physical conditions around $r\simeq5\ r_{\rm g}$
\citep{EHT8}: $n_{\rm e} \simeq10^4-10^7 {\rm \ cm}^{-3}$, $T_{\rm e}\simeq(1-12)×10^{10}$ K. 
The magnetic field strength in the model (50–100G) is slightly larger than, but comparable to, the EHT-inferred value ($\simeq 1-30$ G).
The plasmas inside ISCO have finite values in this paper, while it is assumed to be a vacuum in \citet{Kawashima2019}. 
$F(t)$ is a time-variable function explained later.

We set the fluid four velocity, $u^\mu$, as a mixture of free-fall velocity and Keplerian rotation velocity (e.g., \citet{Pu2016}). The $r$-component of the velocity is
\begin{align} \label{fluid2}
    u^r = u^r_{\rm K} + (1 - C_1) (u^r_{\rm ff} - u^r_{\rm K}),
\end{align}
where $u^r_{\rm K}$ and $u^r_{\rm ff}$ are
$r$-component of the Keplerian and free-fall velocities,
\begin{align} \label{fluid3}
    u^r_{\rm K} &= 
    \left\{
    \begin{array}{ll}
    0 \ (r > r_{\rm ISCO}) \\
    -[2/3r_{\rm ISCO}]^{1/2} [r_{\rm ISCO}/r - 1]^3/2 \ (r \leq r_{\rm ISCO})
    \end{array}
    \right.
     ,\\
    u^r_{\rm ff} &= -[2r (r^2+a^2)]^{1/2} \Sigma^{-1} ,
\end{align}
with $r_{\rm ISCO}$ being the radius of the innermost stable circular orbit.
$C_1$ is a free parameter.
The angular velocity is given by
\begin{align} \label{fluid2}
    \Omega = \Omega_{\rm K} + (1 - C_2) (\Omega_{\rm ff} - \Omega_{\rm K}),
\end{align}
where $C_2$ is a free parameter,
$\Omega_{\rm K}$ and $\Omega_{\rm ff}$ are
\begin{align} \label{fluid3}
    \Omega_{\rm K} &= 
    \left\{
    \begin{array}{ll}
    (r^{3/2} + a)^{-1} \ (r > r_{\rm ISCO}) \\
    (\mathcal{L} + a \mathcal{H}) [r^2 + 2r(1 + \mathcal{H})]^{-1} \ (r \leq r_{\rm ISCO})
    \end{array}
    \right.
     ,\\
    \Omega_{\rm ff} &= 2ar A^{-1},
\end{align}
with $\mathcal{L} = (r_{\rm ISCO}^2 - 2 a \sqrt{r_{\rm ISCO}} + a^2)/(\sqrt{r_{\rm ISCO}^3} - 2\sqrt{r_{\rm ISCO}} + a)$ and $\mathcal{H} = (2r - a \mathcal{L})/\Delta$.
$\theta$-component of the velocity, $u^\theta$, is set to be null.
The $t$-component of the four velocity, $u^t$, is obtained from $u^\mu u_\mu = -1$,
where the components of $u^\mu$ are $u^\mu = (u^t, u^r, 0, \Omega u^t)$.
We set $C_1$ and $C_2$ to be $0.5$ and $0.3$
so that the velocity matches that of recent GRMHD simulations \citep[][]{Chael2021, Narayan2022}. 

Using the time-variable function
\begin{align} \label{fluid4}
    F(t) &= 
    \left\{
    \begin{array}{ll}
    f \ (0 \leq t \leq \Delta t_{\rm d}) \\
    1 \ (0 < t, \Delta t_{\rm d} < t)
    \end{array}
    \right.
    ,
\end{align}
the time variation of the electron number density as well as the magnetic field strength are controlled,
where $f$ is the variation factor, $\Delta t_{\rm d}$ is the duration of the increase of the electron number density. 
It should be noted that this time variation mimics the case of the prompt increase of the mass accretion rate near the BH, rather than the appearance of orbiting hot-spot like features.
In this paper, we focus on the effects of the time-dependent photon propagation taking into account the causality of the light rays. More detailed discussion will be given in Section \ref{sec:result}.
We set $t=0$ when the electron number density of the accretion disk starts to increase.
In all cases, the radiation flux at the 230 GHz band becomes almost 0.3 Jy if $f$ is constant at 1.0.
In this paper, the variation factor, $f$, is set to be $f = 2.0$, $2.5$, $3.0$, $3.5$, and $4.0$. 
Therefore, we calculate 15 cases (3 BH spin parameters $\times$ 5 variation factors) in total. 
The accretion disk is SSA-thin except for the innermost region of the disk($r\lesssim 5r_{\rm g}$). 
Then, increases in density and magnetic field strength induce increases in emissivity as well as absorption coefficient. 
Thus, the accretion disk becomes partially SSA-thick at 230 GHz band at $0 \leq t \leq  \Delta t_{\rm d}$. 
We basically set $\Delta t_{\rm d} = 15\ t_{\rm g}$ ($t_{\rm g}=r_{\rm g}$) since the dynamical timescale at the innermost region is around $15\ t_{\rm g}$,
which corresponds to $\simeq 5$ days for M87 \citep{Satapathy2022, Wong2024}. 
For comparison, $\Delta t_{\rm d} = 10\ t_{\rm g}$ and $\Delta t_{\rm d} = 20\ t_{\rm g}$ are also employed.

The value of the variation factor adopted in this work ($f=2.0-4.0$) is thought to be reasonable as compared to observations. The M87 core flux at 230 GHz was reported to be $\simeq 1.0$ Jy in 2012 \citep{Akiyama2015} and $\simeq 0.5$ Jy in 2019 \citep{EHT1}. This fact indicates that the M87 core flux fluctuates, which may be the year-scale variation, by at least a factor of two. 
In our models, the 230 GHz fluxes for $f=2.0$ reach 2.1 times the baseline value for $a=0.998$ and 2.3 times for $a=0.5$ and $0.75$. The ratio of the peak flux to the baseline value increases with $f$ and for $f=4.0$ it is about 4 (with a maximum of 4.1 for $a=0.5$).
Therefore, the results are consistent with the observations when the variation factor is around 2.0, and in cases of larger variability, a larger variation factor may become more appropriate.
Regarding $\Delta t_{\rm d}$, the value of $15\ t_{\rm g}$ is thought to be reasonable. ALMA observations of SgrA$^*$ \citep{Iwata2020} and GRMHD simulations \citep{Chan2015_2} report that the 230 GHz flux varies on timescales comparable to the dynamical timescale near the innermost region of the accretion flow ($\gtrsim$ $10 t_{\rm g}$). Since both M87 and SgrA$^*$ are LLAGNs, if M87 is also assumed to vary on such a dynamical timescale,
the typical variability timescale 
is expected to be $\simeq 10-20\ t_{\rm g}$, which is consistent with the value adopted here ($\Delta t_{\rm d}=15\ t_{\rm g}$).

\section{Results and Discussion} \label{sec:result}
\subsection{Time variation of simulated images} \label{sec:timevar}
\begin{figure*}[]
  \begin{tabular}{cc}
    \begin{minipage}{\hsize}
      \begin{center}
        \includegraphics[width=0.85\linewidth]{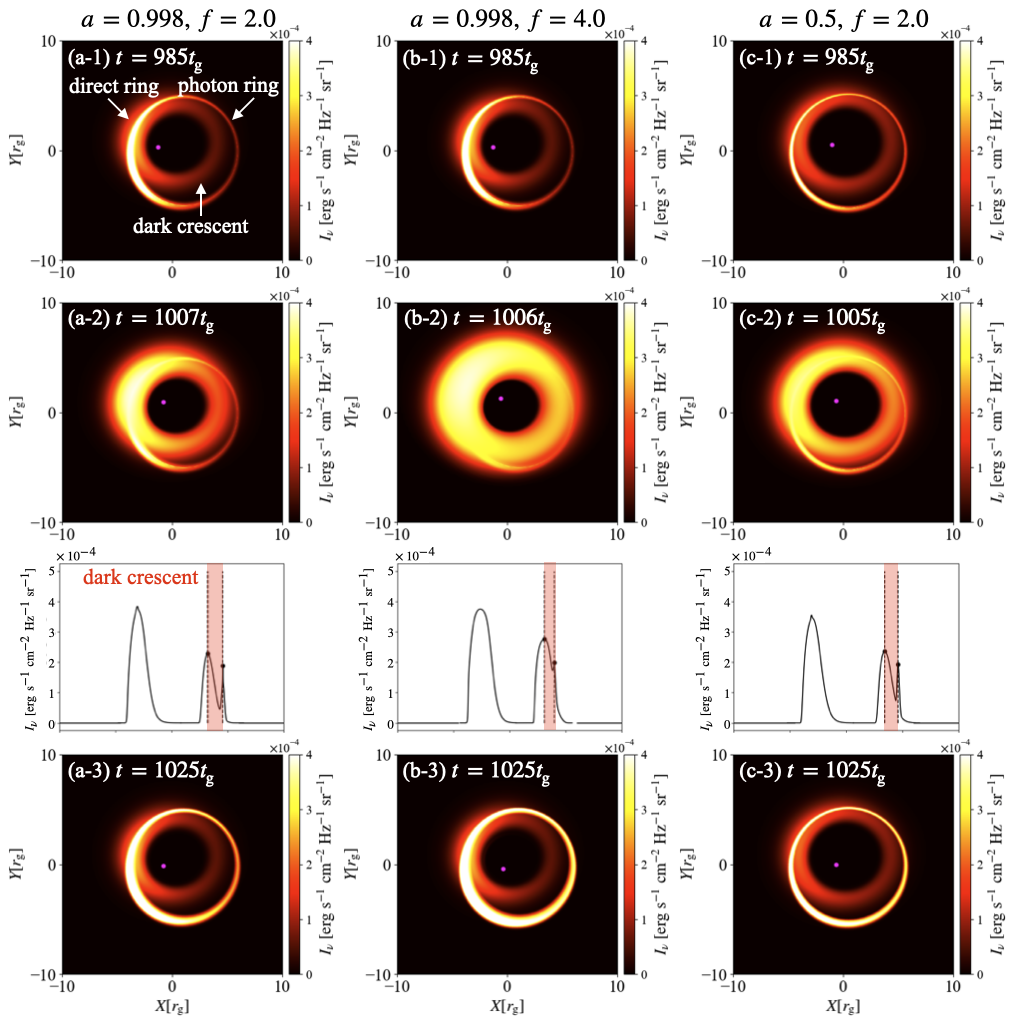}
         \caption{Linear scale time-variable images with $a=0.998, f=2.0$ (left column), $a=0.998, f=4.0$ (middle column) and $a=0.5, f=2.0$ (right column). From the top row to the bottom row, the simulated images at the initial state, at the brightest state and at the time when the photon ring becomes the brightest are shown. The bottom panels of (a-2), (b-2), and (c-2) show the sliced intensity profile at the same time as (a-2), (b-2), and (c-2), which is used to define the dark crescent denoted by red-shaded regions. Magenta circles represent the position of the intensity-weighted center. {Alt text: Intensity maps with three parameters. }}
         \label{fig:intensitymap}
       \end{center}
     \end{minipage}
   \end{tabular}
\end{figure*}

The first, second, and fourth rows of Figure \ref{fig:intensitymap} show the time-variation of the simulated images. The left, middle, and right columns denote results in the case of $a=0.998, f=2.0$, $a=0.998, f=4.0$, and $a=0.5, f=2.0$.
Two kinds of rings appear in the resultant images (see panel (a-1) in Figure \ref{fig:intensitymap}). The ring whose center is located in the bottom right is called the "photon ring". In this paper, the photon ring is defined as the region in which a number of times the geodesics cross the equatorial plane is more than once (so called $n\geq1$ subring).
On the other hand, the ring whose center is located in the upper left is called the "direct ring". The direct ring is defined as the region in which a number of times the geodesics cross the equatorial plane is less than twice (so called $n=0$ subring).
Thus, it should be noted that the region defined as each ring
is completely determined by the BH spin parameter and independent of the accretion flow structure because a number of times the geodesics cross the equatorial plane depends only on the BH spin parameter if the inclination angle is fixed. 
The dark region appearing between the two rings is called the "dark crescent" (hereafter DC, denoted in panel (a-1) of Figure \ref{fig:intensitymap}). The DC also changes with time.
Further, the magenta circles near the center denote the intensity-weighted centers. The position of the intensity-weighted center also changes with time. The time variation of the DC and the intensity-weighted center are discussed later in Section \ref{sec:dc} and Section \ref{sec:intcenter}.

\begin{figure*}[]
  \begin{tabular}{cc}
    \begin{minipage}{\hsize}
      \begin{center}
        \includegraphics[width=0.6\columnwidth]{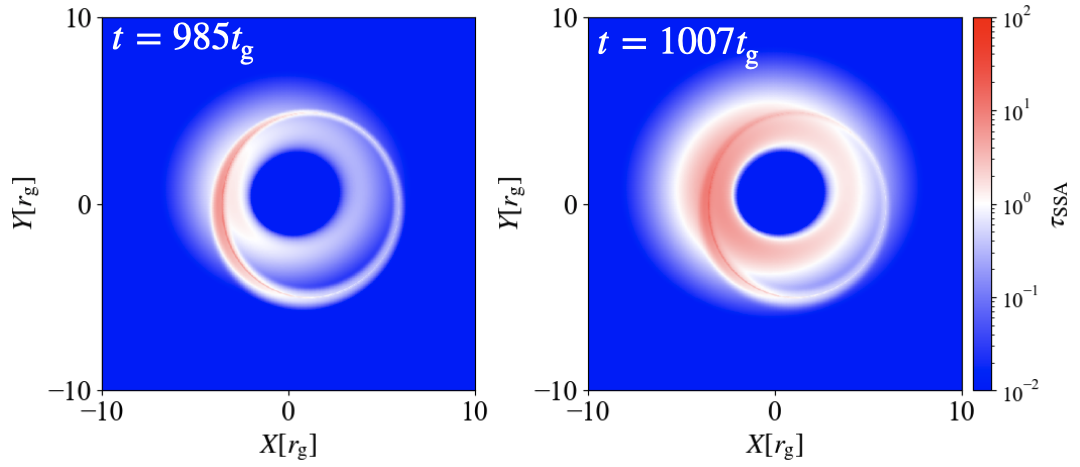}
         \caption{Log-scale map of the optical depth for the synchrotron absorption, $\tau_{\rm SSA}$ in the case of $a=0.998, f=2.0$. The left panel shows the map at the initial state and the right panel shows the map at the brightest state. When the accretion disk emissivity increases, the optically thick region (white to red region) in the direct ring spreads. {Alt text: Optical depth maps at initial and flaring states. }}
         \label{fig:taumap}
       \end{center}
     \end{minipage}
   \end{tabular}
\end{figure*}

After starting the simulations, the moderately bright region in the direct ring (e.g., the domain where the intensity exceeds \(1.0 \times 10^{-4}\ {\rm erg\ s^{-1}\ cm^{-2}\ Hz^{-1}\ sr^{-1}}\)) expands as shown in panels (a-2), (b-2), and (c-2) in Figure \ref{fig:intensitymap}.
This is because the emissivity of the accretion disk increases. 
Indeed, in Figure \ref{fig:taumap}, the optically thick region (white to red area) expands. 
This indicates that the area of the bright region, in which the intensity is almost initially saturated at the blackbody one, becomes large as the disk emissivity increases.
Subsequently, the moderately bright region in the photon ring expands later than that of the direct ring because the photons generating the photon ring reach the observer after orbiting around the BH (see panels (a-3), (b-3), and (c-3) in Figure \ref{fig:intensitymap}). Additionally, the moderately bright region in the direct ring begins to shrink at the same time that the moderately bright region in the photon ring expands. Eventually, the photon ring becomes darker and returns to a steady state at \(t=985t_{\rm g}\).

\begin{figure}[]
  \begin{tabular}{cc}
    \begin{minipage}{\hsize}
      \begin{center}
        \includegraphics[width=\columnwidth]{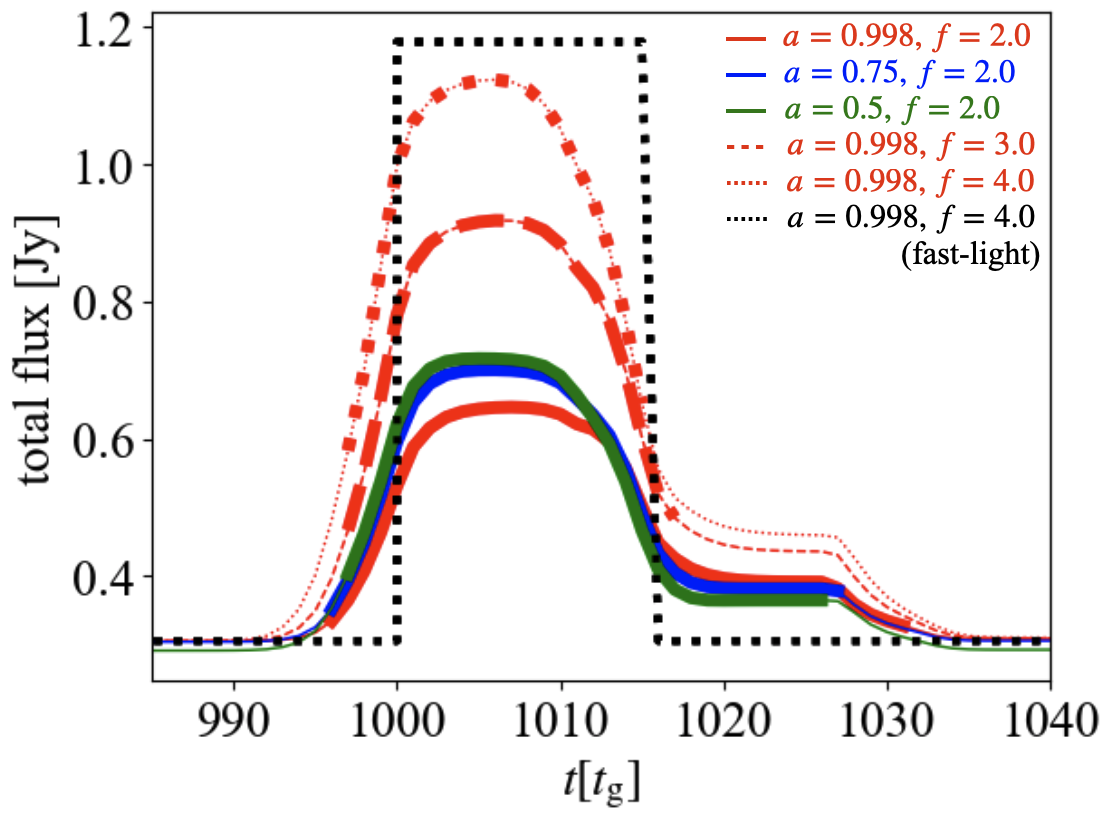}
         \caption{Time variation of the total flux of the images. Red, blue, and green lines denote the results of $a=0.998, 0.75$, and $0.5$, respectively. Solid, dashed, and dotted lines denote the results of $f=2.0, 3.0$, and $4.0$. Furthermore, the black dotted line denotes the total flux calculated with the fast-light approximation in the case of $a=0.998, f=4.0$. {Alt text: Line graph with six lines. }}
         \label{fig:lightcurve}
       \end{center}
     \end{minipage}
   \end{tabular}
\end{figure}

Figure \ref{fig:lightcurve} shows the time variation of the total flux calculated as
\begin{align} \label{Stot}
    S_{\rm tot} (t) = \frac{1}{D^2} \int_{\rm screen} I_\nu (X,Y,t) dXdY,
\end{align}
where $I_\nu (X,Y,t)$ is the intensity at $(X, Y)$, and time $t$.
In $t\lesssim 990t_{\rm g}$, the photons emitted during the low emissivity period (i.e., $f=1.0$) 
reach the screen. Thus, the total flux is nearly 0.3 Jy. The high flux state ($t\simeq 995t_{\rm g}-1010t_{\rm g}$) shows that the direct ring becomes brighter. 
Then, the total flux slightly becomes higher relative to the initial flux via the increase of the flux of the photon ring ($t\simeq 1010t_{\rm g}-1025t_{\rm g}$). 
In $t\gtrsim 1025t_{\rm g}$, the total flux returns to $\simeq$0.3 Jy.
In this figure, the thick lines highlight the period in which the total flux exceeds half of the maximum flux. 
We use the data during this period for estimating the BH spin parameter, which is demonstrated in Section \ref{sec:dc}, \ref{sec:intcenter}, and \ref{sec:estimation}.
The start time ($t_1$) and end time ($t_2$) of each case are summarized in Table \ref{tab:t1t2}. 

\begin{table}
 \begin{center}
 \caption{List of $t_1, t_2$}
 \label{tab:t1t2}
 \begin{tabular}{ccc}
  \hline
  Case & $t_1\  [t_{\rm g}]$ & $t_2\  [t_{\rm g}]$\\
  \hline
  $a=0.998,\ f=2.0$ & 987 & 1023 \\
  $a=0.998,\ f=3.0$ & 989 & 1010 \\
  $a=0.998,\ f=4.0$ & 990 & 1009 \\
  $a=0.75,\ f=2.0$ & 988 & 1020 \\
  $a=0.75,\ f=3.0$ & 990 & 1009 \\
  $a=0.75,\ f=4.0$ & 991 & 1009 \\
  $a=0.5,\ f=2.0$ & 989 & 1019 \\
  $a=0.5,\ f=3.0$ & 991 & 1009 \\
  $a=0.5,\ f=4.0$ & 992 & 1009 \\
  \hline
 \end{tabular}
 \end{center}
\end{table}

The black dotted line shows the result calculated using the fast-light approximation for the case of \(a=0.998, f=4.0\). Note that the time is shifted so that the start of the increase in the total flux aligns with \(t=1000t_{\rm g}\). Comparing the red and black dotted lines, it is evident that the fast-light calculation cannot reproduce the gradual increase and decrease of the total flux. Furthermore, in the fast-light calculation, the flux variation of the direct ring and the photon ring occurs simultaneously, leading to an overestimation of the maximum value of the total flux. Additionally, the fast-light calculation cannot replicate the scenario where only the direct ring or the photon ring becomes brighter.
One may find that the difference between our results obtained by the slow-light calculations and fast-light approximation is more remarkable than those demonstrated by, e.g., \citet{Dexter2010}, since we focus on a rapid and prominent increase of the emissivity of the accretion flow. We emphasize that our slow-light calculations are more essential to consider the time variation of images demonstrated in Section \ref{sec:timevar} and trajectories of the intensity-weighted center, which will be shown in Section \ref{sec:intcenter}.

\subsection{Width of the dark crescent} \label{sec:dc}
In order to calculate the width of the DC, we determine the image center and draw lines radially from the image center. We define the width of the DC as an interval between the positions of the peak intensities of the direct ring and the photon ring on each line. The details are as follows.

Firstly, we calculate the image center. In this paper, we fix the image center to the intensity-weighted center at $t=t_{\rm max}$, $(X_c(t_{\rm max}),\ Y_c(t_{\rm max}))$, where $t_{\rm max}$ is the time that $S_{\rm tot}(t)$ reaches the maximum. The position of the intensity-weighted center at $t$, $(X_c(t),\ Y_c(t))$, can be calculated by
\begin{align} \label{intcenter}
\begin{aligned}
    X_c(t) = \frac{\int_{\rm screen} I_\nu (X,Y,t) X dX dY}{\int_{\rm screen} I_\nu (X,Y,t) dX dY}, \\
    Y_c(t) = \frac{\int_{\rm screen} I_\nu (X,Y,t) Y dX dY}{\int_{\rm screen} I_\nu (X,Y,t) dX dY}.
\end{aligned}
\end{align}
Next, we calculate the distribution of the intensity along a line passing through $(X_c(t_{\rm max}),\ Y_c(t_{\rm max}))$, and compute the interval between the positions of the peak intensity in the direct ring region and the photon ring region in $X>X_c(t_{\rm max})$ on the observer's screen. We calculate these intervals using numerous lines passing through $(X_c(t_{\rm max}),\ Y_c(t_{\rm max}))$, and set the maximum value of intervals as the width of the DC, $W_{\rm DC}(t)$. We note that if the two peaks do not appear, the line is excluded from consideration.
For example, we show the sliced intensity distribution along the line used to determine the DC width in the third row of Figure \ref{fig:intensitymap}. The red shaded region represents the DC. It is important to note that the DC width calculated in this manner is independent of how $X$ and $Y$ axes are set on the screen. 

\begin{figure}[]
  \begin{tabular}{cc}
    \begin{minipage}{\hsize}
      \begin{center}
        \includegraphics[width=\columnwidth]{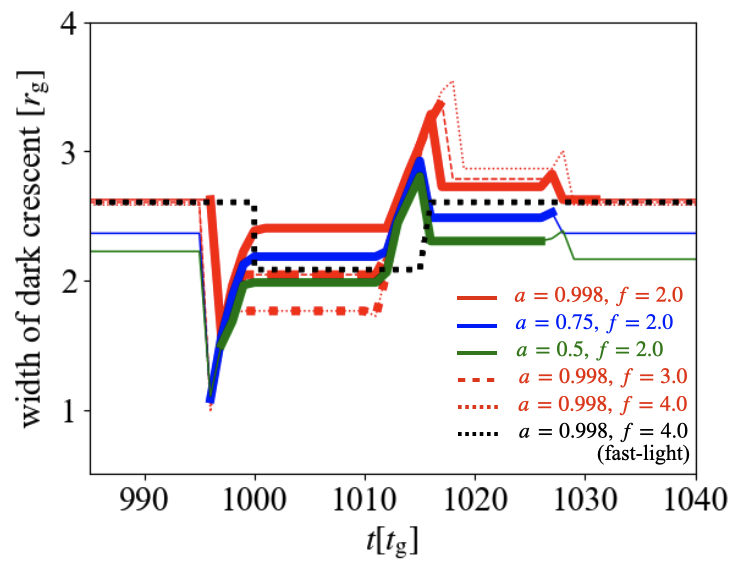}
         \caption{Time variation of the width of the DC. Red, blue, and green lines denote the results of $a=0.998, 0.75$, and $0.5$, respectively. Solid, dashed, and dotted lines denote the results of $f=2.0, 3.0$, and $4.0$. The black dotted line represents the case calculated by the fast-light calculation with $a=0.998, f=4.0$. {Alt text: Line graph with six lines. }}
         \label{fig:var_dcw}
       \end{center}
     \end{minipage}
   \end{tabular}
\end{figure}

Figure \ref{fig:var_dcw} shows the time variation of the width of the DC. As in Figure \ref{fig:lightcurve}, the period between $t_1$ and $t_2$ is shown by the thick line. We can see the DC width in $1000\ t_{\rm g} \lesssim t \lesssim 1010\ t_{\rm g}$ is smaller than that in $t \lesssim 995t_{\rm g}$. 
This is because the position of the peak intensity generating the left edge of the DC (left edge of the red region in the third row of Figure \ref{fig:intensitymap}) moves away from the center of the screen after increasing the emissivity of the accretion disk.

The peak-intensity position on the screen is determined by the special relativistic motion of the fluid and the gravitational redshift at the photosphere on the geodesic.
To understand the time variation of the peak-intensity position on the direct ring, we compare the geodesic reached $(X,Y)=(3.75 r_{\rm g},0)$ which corresponds to the peak-intensity position prior to the increase in the disk emissivity (geodesic A) and the geodesic reached $(X,Y)=(4.76 r_{\rm g},0)$ which corresponds to the peak-intensity position after increasing the disk emissivity (geodesic B) as an example. Before increasing the disk emissivity, the intensity on geodesic A is higher than that of geodesic B because geodesic A passes through the region closer to the BH where the electron temperature and magnetic field strength are higher. The intensity on the geodesic B becomes higher than that of the geodesic A after increasing the disk emissivity due to the following reason. As increasing the disk emissivity, the optical depth along the geodesic A and B becomes higher (7.25 and 4.36 along the geodesic A and B, respectively), and the photosphere appears at $(r,\theta)\simeq(2.6 r_{\rm g}, 0.47 \pi)$ and $(3.6 r_{\rm g}, 0.48 \pi)$ along the geodesic A and B, respectively. The electron temperature at the photosphere is $T_{\rm e}=5.6 \times 10^{10} {\rm K}$ for the geodesic A and $4.1 \times 10^{10} {\rm K}$ for geodesic B, and the redshift factor $g$ corresponding to the ratio of the frequency between the fluid rest frame and the Boyer-Lindquist coordinates is $0.26$ for the geodesic A and $0.41$ for the geodesic B. The redshift factor of the geodesic B is greater than that of the geodesic A because both $|u^r|$ and the effect of the gravitational redshift becomes smaller farther from the BH. Although the electron temperature at the photosphere on the geodesic A is higher than that of the geodesic B, the intensity integrated along the geodesic B becomes higher than geodesic A because of a smaller redshift factor.
Thus, the peak intensity generating the left edge of the DC moves away from the center of the screen.
Around \(t \simeq 1015-1025\ t_{\rm g}\), the flux and width of the direct ring return to their initial state. Meanwhile, the photon ring becomes brighter due to the increased emissivity with a time delay, causing the position of the peak intensity of the photon ring to shift slightly in the bottom right direction. Consequently, the DC width also increases significantly. Afterward, at \(t \gtrsim 1030\ t_{\rm g}\), the photon ring and the DC width revert to their initial states. These time variations become remarkable as the variation factor increases. Thus, with a larger variation factor, the DC width is smaller in \(1000\ t_{\rm g} \lesssim t \lesssim 1010\ t_{\rm g}\) and larger around \(t \simeq 1015-1025\ t_{\rm g}\) (see the red solid, dashed, and dotted lines). Additionally, the DC width increases significantly around \(t \simeq 1015\ t_{\rm g}\) and decreases around \(t \simeq 995\ t_{\rm g}\). This occurs because an increase (decrease) in the variation factor causes the photosphere seen from the observer to move further from (closer to) the BH so the physical quantities at the photosphere vary, leading to significant fluctuations in the position of the peak intensity in the direct ring region during the changes in total flux.

Comparing the red, blue, and green solid lines, we find that the DC width becomes large when the BH spin parameter is high. This is because, with a higher BH spin parameter, the photon ring shifts more in the right direction, although the position of the direct ring does not depend much on the BH spin parameter (see also panels (a-2) and (c-2) in Figure \ref{fig:intensitymap} and Section 4 in TK19). Therefore, the width of the DC becomes larger when the BH spin parameter is higher.
We can see the spikes at $t \simeq 995 t_{\rm g}$ and $1015 t_{\rm g}$. This is because the multiple local maxima of the intensity temporarily appear in the DC region at the beginning of the increase (decrease) in the flux of the direct ring. These spikes seen at $t \simeq 995 t_{\rm g}$ and $1015 t_{\rm g}$ have little effect on the time-averaged width of the DC mentioned later.

The black dotted line shows the result with the fast-light approximation in the case of $a=0.998,f=4.0$. By comparing this line with the red dotted line, the DC width is not calculated correctly in the fast-light calculation. Furthermore, the DC width is overestimated in $1000\ t_{\rm g} \lesssim t \lesssim 1010\ t_{\rm g}$.

\subsection{Trajectory of the intensity-weighted center} \label{sec:intcenter}
\begin{figure}[]
  \begin{tabular}{cc}
    \begin{minipage}{\hsize}
      \begin{center}
        \includegraphics[width=\columnwidth]{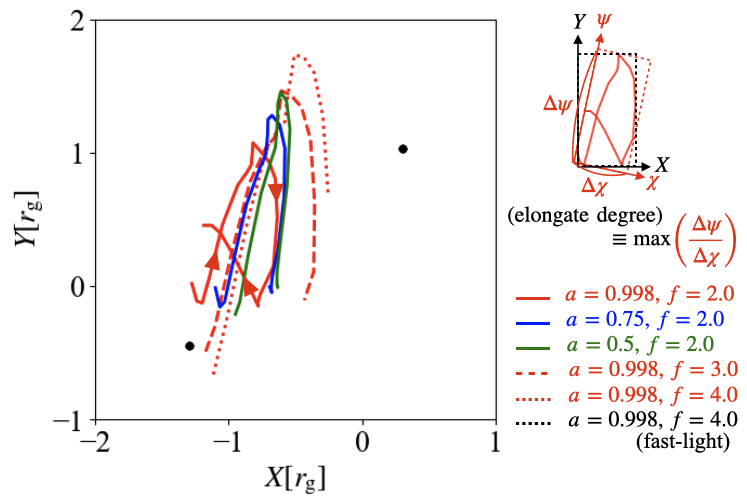}
         \caption{Trajectories of the intensity-weighted center on the screen. Red, blue, and green lines denote the results of $a=0.998, 0.75$, and $0.5$, respectively. Solid, dashed, and dotted lines denote the results of $f=2.0, 3.0$, and $4.0$. The black dots are the position of the intensity-weighted center calculated by the fast-light calculation with $a=0.998, f=4.0$. The bottom left and top right dots correspond to the initial state and flaring state. The top right schematic illustration shows the definition of the $\chi,\psi$ axis and the elongate degree of the trajectory. {Alt text: Trajectories of image center with six parameters. }}
         \label{fig:inttr}
       \end{center}
     \end{minipage}
   \end{tabular}
\end{figure}
Figure \ref{fig:inttr} shows the trajectories of the intensity-weighted center for each model. 
It can be seen that the intensity-weighted center on the screen first moves to the upper right, then to the bottom, and finally to the upper left in the case of $a=0.998, f=2.0$ (red solid line). The slight shift to the bottom right is mentioned later. This can be understood that (i) the intensity at the top of the screen changes slightly later than at the bottom due to the inclination angle, (ii) the increase in intensity caused by the increase in emissivity is remarkable on the right side than on the left side of the screen, and (iii) the flux of the photon ring changes later than that of the direct ring.

\begin{figure}[]
  \begin{tabular}{cc}
    \begin{minipage}{0.94\hsize}
      \begin{center}
        \includegraphics[width=\columnwidth]{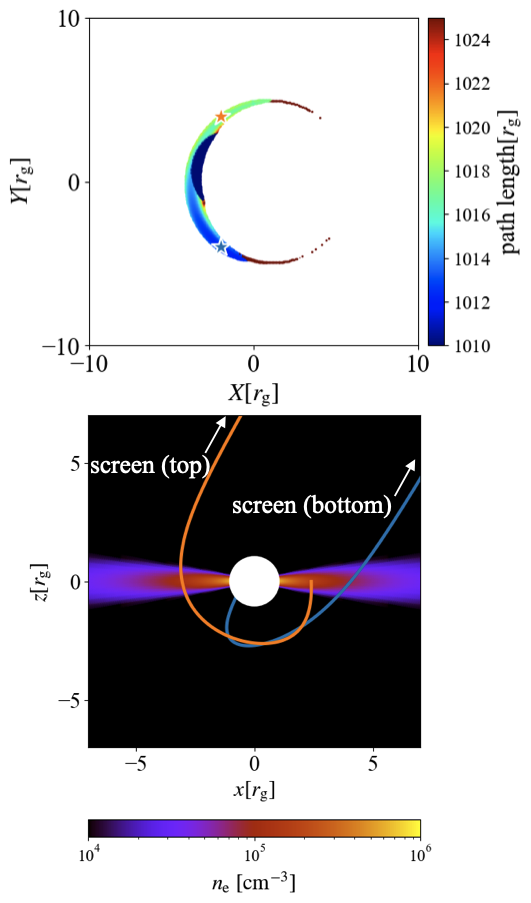}
         \caption{The top panel represents the map of the path length of photons emitted from the photosphere to the screen in the photon ring. The bottom panel shows two examples of geodesics generating the photon ring. The color contour denotes the electron number density at the initial state. The central white filled circle represents the BH. Orange and blue stars in the top panel denote the end-points of two geodesics shown in the bottom panel. {Alt text: Path length map and two geodesics.} }
         \label{fig:pathlength}
       \end{center}
     \end{minipage}
   \end{tabular}
\end{figure}

The reason why the intensity-weighted center moves to the top (bottom) with the increase (decrease) in the emissivity can be understood by considering the difference in the photon path lengths that reach the top and bottom of the screen.
In the direct ring, the path length of the photon emitted from the closer to the screen is shorter than that from the farther from the screen. This difference is obviously determined by the inclination angle.
On the other hand, it is not obvious whether the difference in the photon path length in the photon ring is determined by the same reason as that in the direct ring because the photon ring is generated by the photons orbiting around the BH. The top panel of Figure \ref{fig:pathlength} shows the photon path length from the photosphere to the screen in the photon ring.
In the top panel of Figure \ref{fig:pathlength}, we plot the region with $\tau_{\rm SSA} \geq 1$ because the flux of the region with $\tau_{\rm SSA}<1$ is only 30\% of the total flux in the photon ring. The green region (i.e., path length $\simeq 1017 r_{\rm g}$) on the top and the blue region (i.e., path length $\simeq 1012 r_{\rm g}$) on the bottom are distributed, indicating that the difference in the path length between the top and bottom of the photon ring is approximately $5 r_{\rm g}$. This difference is also determined by the inclination angle. 

The bottom panel of Figure \ref{fig:pathlength} shows examples of the geodesic reach the top of the screen (orange line) and the bottom of the screen (blue line). Positions at which each geodesic reaches the screen are denoted by orange and blue stars in the top panel of Figure \ref{fig:pathlength}.
As seen from the bottom panel of Figure \ref{fig:pathlength}, we can find that the geodesics in the region on the observer's side relative to the equatorial plane ($z>0$) are not very curved. Using this and the fact that the diameter of the photon ring is approximately $10 r_{\rm g}$, the path length of the blue geodesic is approximately $10r_{\rm g} \times \cos (30^\circ)=5r_{\rm g}$ shorter than that of the orange geodesic.
In contrast, the geodesics in the region on the opposite side of the observer relative to the equatorial plane ($z<0$) are significantly curved. We found that the difference in the photon path lengths of two geodesics shown in the bottom panel of Figure \ref{fig:pathlength} in $z<0$ is almost $0.5r_{\rm g}$, and it is smaller enough than the difference in the photon path length in $z>0$. The differences in the photon path length for other geodesics are also similar.

We can, therefore, conclude that the intensity-weighted center moves top and bottom because the top of the photon ring changes later than the bottom with approximately $5t_{\rm g}$. We note that the path length of the dark blue region on the left side in Figure \ref{fig:pathlength} becomes short because the optical depth becomes greater than 1 at the first crossing of the equatorial plane seen from the observer. Furthermore, the path length of the thin red region on the right side becomes longer because the photon crosses the equatorial plane more than three times until $\tau \geq 1$. However, the change in the intensity in this region is sufficiently small and the influence contributing to the shift of the intensity-weighted center is negligible. 

Next, we compare the photon ring in the initial state and bright state. As shown in Figure \ref{fig:taumap}, the optically thick region on the left side is initially narrow (around $X\simeq -4 r_{\rm g}, Y\simeq 0$), but as disk emissivity increases, it widens and the left half of the ring becomes optically thick. The optical depth in the right half of the photon ring also becomes higher. Although intensity tends to rise with an increase in optical depth, in optically thick conditions, intensity becomes less dependent on optical depth because that is saturated at the blackbody intensity. 
Therefore, the increase in intensity caused by higher emissivity is greater on the right side of the photon ring than on the left (see also Section \ref{sec:timevar} and \ref{sec:dc}). 
On the left side of the photon ring, although the optical depth in the area whose optical depth is initially around 1 increases significantly, the optically thick region is not expanded much. On the right side of the photon ring, the optical depth increases from less than 1 to greater than 1 over a larger area. This is the reason why the increase in the intensity caused by the increase in the disk emissivity on the right side of the photon ring is higher. 

Regarding the direct ring, the increase in intensity on the right side also becomes greater. Consequently, the flux increase basically makes the intensity-weighted center shift toward the right.

As we additionally consider the time delay between the photon ring and the direct ring (see Section \ref{sec:timevar}), 
we summarize the trajectory of the intensity-weighted center changes in the following steps:
\begin{enumerate}
    \item As emissivity increases, the bottom of the direct ring first becomes brighter, followed by the gradual brightening of the upper side. Finally, the whole of the direct ring becomes brighter. Because of the effects of the increase in the emissivity (the intensity-weighted center moves to the right) and gradual upper side brightening (the intensity-weighted center moves to the upper), the intensity-weighted center moves to the upper right.
    \item The direct ring begins to darken from the bottom, and almost simultaneously, the photon ring starts to brighten from the bottom. Over time, the entire direct ring becomes dark, while the photon ring becomes brighter overall. During this process, because the effect of the photon ring brightening (which shifts the intensity-weighted center downward) is greater than the effect of the direct ring darkening (which shifts the intensity-weighted center upward), the intensity-weighted center shifts downward. More precisely, the intensity-weighted center first moves downward and to the right, and then moves downward and to the left. In other words, the trajectory of the intensity-weighted center curves outward to the right. This is because in the first half, the shift to the right due to the brightening of the photon ring dominates, while in the latter half, the shift to the left due to the darkening of the direct ring dominates.
    \item Finally, the photon ring gradually becomes darker from the bottom to the top. The intensity-weighted center moves to the upper left.
\end{enumerate}
In Figure \ref{fig:inttr}, the trajectories of the intensity-weighted center are not closed because we use the limited period mentioned in Section \ref{sec:timevar}.

We note that the intensity-weighted center slightly shifts to the bottom right ($X\simeq -1.0r_{\rm g}$) at the beginning of the increase in disk emissivity in the cases of $a=0.998, f=2.0$ and $a=0.75, f=2.0$ in Figure \ref{fig:inttr}. 
This occurs at the moment the flux in the direct ring begins to brighten. In cases other than $a=0.998, f=2.0$ and $a=0.75, f=2.0$, this moment is not contained in $t_1 \leq t \leq t_2$, so that slight shift of the intensity-weighted center to the bottom right cannot be seen in Figure \ref{fig:inttr} in these cases.
In addition, the shift of the intensity-weighted center to the upper left cannot be seen because the effect of itemization 3 in the cases other than $a=0.998, f=2.0$ occurs in $t \geq t_2$.
Basically, such a result can be understood that the asymmetry in the $X$ direction of the images becomes remarkable when the BH spin parameter is high because of the photon trajectory and Doppler effect. Then, the motion of the intensity-weighted center in the $X$ direction is suppressed when the BH spin parameter is lower. In addition, the variation of the intensity-weighted center in the $Y$ direction becomes remarkable when the variation factor is large. This is because a larger variation factor results in a larger photosphere and the direct ring spreads significantly in the $Y$ direction when the variation factor is large.
As a result, the trajectories of the intensity-weighted center become more elongated when the BH spin parameter is lower and/or the variation factor is larger.

The black dots denote the positions of the intensity-weighted center obtained by the fast-light calculation in the case of $a=0.998,f=4.0$. The trajectory of the intensity-weighted center can be calculated only when we perform the slow-light calculation.

\subsection{Black hole spin estimation} \label{sec:estimation}
We suggest a novel method to estimate the BH spin parameter using the time-averaged DC width normalized by the gravitational radius,
\begin{align}
    \langle w_{\rm DC} \rangle = \left( \frac{1}{t_2-t_1} \int^{t_2}_{t_1} W_{\rm DC}(t) dt \right) \frac{1}{r_{\rm g}},
\end{align}
and the "elongate degree" that represents the significance of the elongation of the trajectory of the intensity-weighted center.
The elongate degree is defined by the following procedure.
First, we consider the orthogonal coordinate ($\chi,\psi$) as shown in top right panel of Figure \ref{fig:inttr}.
The axes of $\chi$ and $\psi$ are set in such a way that $\Delta \psi / \Delta \chi$ becomes maximum, where $\Delta \chi $ ($ \Delta \psi$) denotes the difference between the maximum and minimum values of  $\chi$ ($\psi$) along the trajectories of intensity-weighted centers.
Finally, we define the elongate degree as the maximum value of $\Delta \psi / \Delta \chi$ as follows:
\begin{align}
    {\rm (elongate\,\,\,degree)} 
    =   \frac{\Delta \psi}{\Delta \chi}.
\end{align}

\begin{figure}[]
  \begin{tabular}{cc}
    \begin{minipage}{\hsize}
      \begin{center}
        \includegraphics[width=\columnwidth]{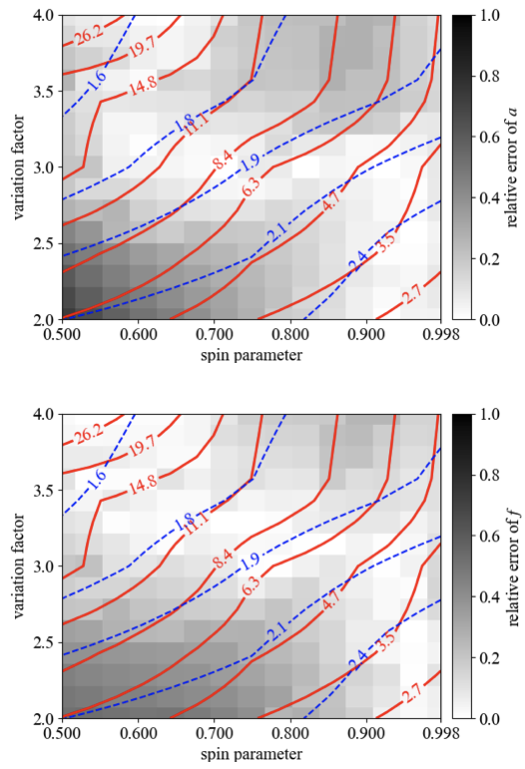}
         \caption{Contours of the time-averaged DC width and the elongate degree on the spin parameter--variation factor plane. Red solid lines and blue dashed lines denote the contours of the elongate degree and the time-averaged DC width, respectively. These contours are plotted by linear interpolating from the results of 15 models. The grayscale background shows the relative error of the estimated spin parameter (top panel) and variation factor (bottom panel) assuming that the time-averaged DC width and the elongate degree are determined with 10\% error. {Alt text: Contours of dark crescent width and elongate degree.}
         }
         \label{fig:afcontour}
       \end{center}
     \end{minipage}
   \end{tabular}
\end{figure}

Figure \ref{fig:afcontour} shows the distribution of $\langle w_{\rm DC} \rangle$ (blue dashed lines) and the elongate degree (red solid lines) on the plane of the BH spin parameter and variation factor. This distribution is obtained by the linear interpolation of the simulated results of 15 models. 
The grayscale background in Figure \ref{fig:afcontour} shows the relative error of the spin parameter (top panel) and the variation factor (bottom panel) assuming that the time-averaged DC width and the elongate degree are determined with an accuracy of 10\%. The closer to white, the more accurately the spin parameters can be determined. The white to light gray regions appear from the bottom right to the top left. Thus, the spin parameter and the variation factor can be estimated accurately in these regions with the relative error of $10-20$ \%.
For example, if $\langle w_{\mathrm{DC}} \rangle = 2.1$ and the elongate degree $=4.7$ are determined with an accuracy of 10\%, the spin parameter and variation factor can be estimated as $a \simeq 0.88$ and $f \simeq 2.8$, respectively, with an accuracy of 10--20\%.
On the other hand, the relative errors in the top right and bottom left become large. In these regions, the relative error of the spin parameter is greater than $\simeq 20$ \%. 
In Figure \ref{fig:afcontour}, the mean relative errors of the spin parameter and the variation factor are 16\% and 15\%, respectively.

The relative errors of the spin parameter and the variation factor tend to be large where two sets of contour lines are nearly parallel and the interval of contour lines is wider, because the area on the spin parameter--variation factor plane where $\langle w_{\rm DC} \rangle$ and elongate degree match within an accuracy of 10\% becomes larger. In addition, $\langle w_{\rm DC} \rangle$ contours and elongate degree contours may have more than one intersection. In this 
situation, the relative errors of the spin parameter and the variation factor are also expected to be large, because 
multiple solutions for the spin parameter and variation factor can be estimated from the same combination of $\langle w_{\rm DC} \rangle$ and elongate degree. To resolve these issues, it may be necessary to combine our method with other approaches. 

\begin{figure}[]
  \begin{tabular}{cc}
    \begin{minipage}{\hsize}
      \begin{center}
        \includegraphics[width=\columnwidth]{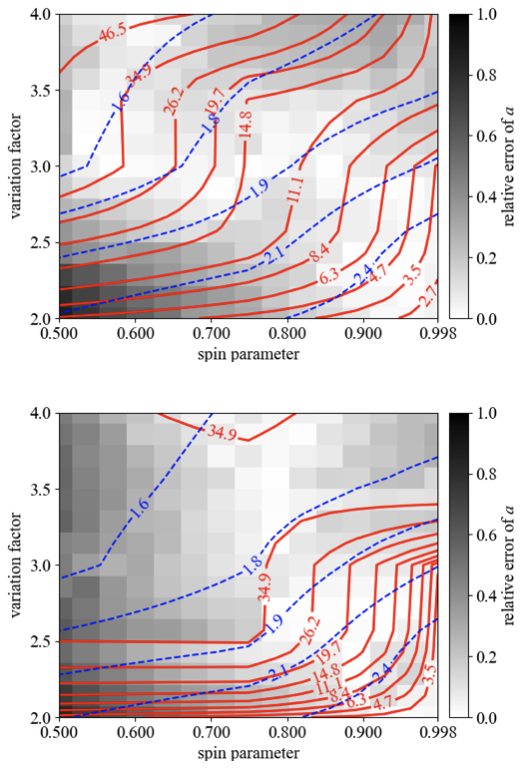}
         \caption{Same contours as Figure \ref{fig:afcontour}, but the observation cadence is $2\ t_{\rm g}$ (top panel) and $4\ t_{\rm g}$ (bottom panel). {Alt text: Contours of dark crescent width and elongate degree.}}
         \label{fig:afcontourdt}
       \end{center}
     \end{minipage}
   \end{tabular}
\end{figure}

Regarding the observing cadence, little difference is found between a cadence of 1 $t_{\rm g}$ ($\simeq$ 0.5 day) and 2 $t_{\rm g}$ ($\simeq$ 1 day), whereas a cadence of 4 $t_{\rm g}$ ($\simeq$ 2 days) leads to a decrease in the accuracy of the spin measurement. This trend is evident with comparing the top panel of Figure \ref{fig:afcontour} and Figure \ref{fig:afcontourdt}.
The top (bottom) panel of Figure \ref{fig:afcontourdt} shows the result with the observation cadence of $2\ t_{\rm g}$ ($4\ t_{\rm g}$) by time-averaging two (four) simulated images generated every $1\  t_{\rm g}$. For example, in the cases with the observation cadence of $1\ t_{\rm g}$ and $2\ t_{\rm g}$, near the parameter combinations with relatively high 
accuracy ($a=0.5,f=4.0$ and $a=0.998,f=2.0$), the relative error is $\simeq$ 10\%. In contrast, when the cadence is 4 $t_{\rm g}$, the area of the white region becomes smaller, and 
the white region still remains in the bottom right region. Therefore, except for cases of $a=0.998, f=2.0$, a cadence of less than about 2 $t_{\rm g}$ is desirable.

\begin{figure}[]
  \begin{tabular}{cc}
    \begin{minipage}{\hsize}
      \begin{center}
        \includegraphics[width=\columnwidth]{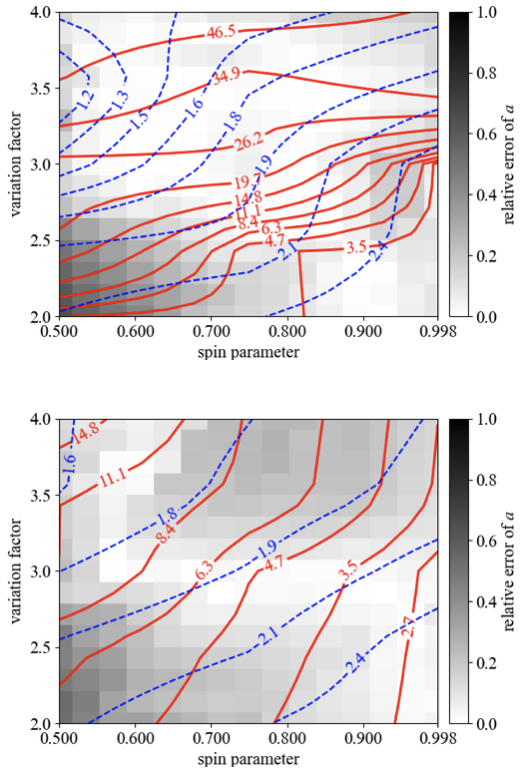}
         \caption{Same contours as Figure \ref{fig:afcontour}, but the duration of the time variation of the accretion disk is $\Delta t_{\rm d}=10\ t_{\rm g}$ (top panel) and $\Delta t_{\rm d}=20\ t_{\rm g}$ (bottom panel). {Alt text: Contours of dark crescent width and elongate degree.}}
         \label{fig:afcontour1020}
       \end{center}
     \end{minipage}
   \end{tabular}
\end{figure}

In the discussion so far, we fixed $\Delta t_{\rm d}=15t_{\rm g}$. In order to study the dependence on $\Delta t_{\rm d}$, we show the results in the case of $\Delta t_{\rm d}=10 t_{\rm g}$ and $20t_{\rm g}$ in Figure \ref{fig:afcontour1020}. As $\Delta t_{\rm d}$ changes, $t_1, t_2$ (see Section \ref{sec:timevar} and Table \ref{tab:t1t2}) are also changed.
By comparing Figure \ref{fig:afcontour} and Figure \ref{fig:afcontour1020}, we can see $\langle w_{\rm DC} \rangle$ shown by blue dashed lines does not significantly depend on $\Delta t_{\rm d}$.
As we can see from Figure \ref{fig:lightcurve} and Figure \ref{fig:var_dcw}, the period in which the DC width significantly changes nearly coincides with the period in which the flux of the direct ring greatly varies. These periods are always included in $t_1 \leq t \leq t_2$ if $\Delta t_{\rm d}$ changes. Thus, the time-averaged DC width $\langle w_{\rm DC} \rangle$ is less affected if $\Delta t_{\rm d}$ changes.

As shown in the top panel of Figure \ref{fig:afcontour1020}, the elongate degree becomes significantly large in the case of $\Delta t_{\rm d}=10 t_{\rm g}$. This is because the trajectory of the intensity-weighted center does not curve outward on the right (see the second itemization of Section \ref{sec:intcenter}). In the case of $\Delta t_{\rm d}=10 t_{\rm g}$, the brightening of the photon ring begins after the darkening of the entire direct ring. Thus, unlike the case of $\Delta t_{\rm d}=15 t_{\rm g}$, there is no situation where the direct ring and the photon ring are brightening simultaneously, even partially.
As a result, the trajectory of the intensity-weighted center does not curve outward on the right ($\Delta \chi$ becomes small), and the elongate degree becomes large. In contrast, in the case of $\Delta t_{\rm d}=20 t_{\rm g}$, the brightening of the photon ring begins before finishing the darkening of the direct ring. A situation arises in which the direct ring and the photon ring are simultaneously brightening.
Therefore, the trajectory of the intensity-weighted center significantly curves outward on the right ($\Delta \chi$ becomes large), and the elongate degree becomes the same as that for $\Delta t_{\rm d} = 15t_{\rm g}$ (see the bottom panel of Figure \ref{fig:afcontour1020}). 
As in Figure \ref{fig:afcontour}, the white to light gray regions appear from the bottom right to the top left, with their area in the case of $\Delta t_{\rm d} = 10\,t_{\rm g}$ (top panel of Figure~9) being slightly wider than that in the case of $\Delta t_{\rm d} = 15\,t_{\rm g}$.
The mean relative errors of the estimated spin parameter are 10\% for $\Delta t_{\rm d} = 10\,t_{\rm g}$ 
and 14\% for $\Delta t_{\rm d} = 20\,t_{\rm g}$, 
respectively. These mean relative errors are nearly the same as that in the case of $\Delta t_{\rm d} = 15\,t_{\rm g}$ (16 \%), and do not change remarkably when $\Delta t_{\rm d}$ differs.



\begin{figure}[]
  \begin{tabular}{cc}
    \begin{minipage}{\hsize}
      \begin{center}
        \includegraphics[width=\columnwidth]{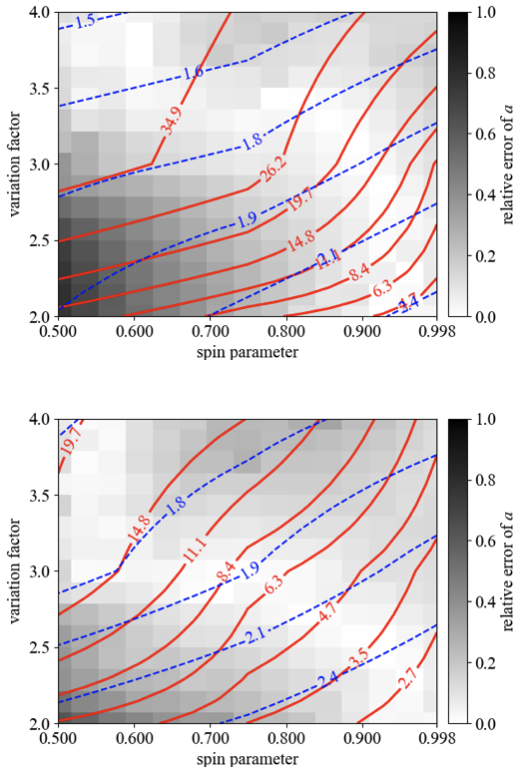}
         \caption{Same contours as Figure \ref{fig:afcontour}, but the electron number density and electron temperature are proportional to $n_{\rm e}\propto r^{-1.1}$ and $T_{\rm e}\propto r^{-0.84}$ (top panel), and $\beta_{\rm mag}= 1.0$ (bottom panel). {Alt text: Contours of dark crescent width and elongate degree.}}
         \label{fig:afcontourfluid}
       \end{center}
     \end{minipage}
   \end{tabular}
\end{figure}

Even when the accretion flow model is slightly different, the estimation accuracy of the spin parameter is not significantly changed. The top panel of Figure \ref{fig:afcontourfluid} shows the result with the radial power-law indices of –1.1 and –0.84 for the electron number density and electron temperature, respectively. The bottom panel of Figure \ref{fig:afcontourfluid} shows the result with $\beta_{\rm mag}=1.0$. The other conditions are identical to the top 
panel of Figure \ref{fig:afcontour}.
A common feature is that the relative error becomes larger in the top right and the bottom left regions (gray regions) and smaller in the bottom right and the top left regions (white to light gray regions) of each panel. The mean relative errors of the spin parameter are 17\% (top panel) and 14\% (bottom panel), respectively, and remain nearly unchanged within the range tested here despite differences in the accretion flow structure.
However, we note that the structure of the accretion flow must be constrained by other observations, since the estimated spin parameter can differ when the accretion flow structure is different, even if the time-averaged DC width and the elongate degree are the same.

\tc{
}

\section{Summary} \label{sec:summary}
In this work, 
we propose a novel method for assessing the BH spin parameter of supermassive BH surrounded by less luminous accretion disks, primarily targeting M87, by considering the situation where the innermost region of the disk temporarily becomes optically thick for synchrotron self-absorption due to a rapid increase in emissivity (absorption coefficient).
Since it is necessary to calculate the time variation of images by taking into account the time delay caused by differences in photon path length, the fast-light approximation cannot be applied. In this study, we use the time-dependent general relativistic radiative transfer code \texttt{CARTOON}.


Due to the difference in photon path lengths, among the two rings appearing on the observer's screen, the direct ring brightens (darkens) first when the emissivity increases (decreases). The flux of the photon ring changes with a delay.
Especially, in the case of the duration that the emissivity increases $\Delta t_{\rm d}=15t_{\rm g}$, the decrease in the flux of 
the direct ring and the increase in that of 
the photon ring occur almost simultaneously. 
%
%

In more detail, differences in photon path length also influence which part of each ring exhibits flux increases or decreases first. 
Since photons from the side of the disk closer to the BH reach the observer's screen first, the flux changes occur on one side rather than across the entire ring. 
Each ring is brighter on the side where photons co-rotate relative to the BH, but the region with a greater increase in brightness due to emissivity enhancement is on the opposite side.
Due to such time variations,
the intensity-weighted center exhibits complex time variations on the observer's screen.
In addition, the brightening of the direct ring reduces the width of the dark crescent (DC) which is the darker region between the photon ring and the direct ring. 

As the amplitude of the BH spin parameter increases and the variation factor decreases, the time-averaged DC width $\langle w_{\rm DC} \rangle$ is larger. Conversely, as the amplitude of the BH spin parameter decreases and the variation factor decreases, the elongate degree, which characterizes the trajectory of the intensity-weighted center of the images, is larger. We showed that the BH spin parameter (and the variation factor) can be estimated by combining the time-averaged DC width, $\langle w_{\rm DC} \rangle$, and the elongate degree. This is a novel idea for estimating 
the BH spin parameter. 
For instance, 
the spin parameter and variation factor can be estimated as $a \simeq 0.88$ and $f \simeq 2.8$, respectively, with an accuracy of 10--20\%, if $\langle w_{\mathrm{DC}} \rangle = 2.1$ and the elongate degree of $4.7$ are determined with an accuracy of 10\%.

It should be noted that the DC width and elongate degree depend on the conditions (e.g., observation cadence, $\Delta t_{\rm d}$, accretion flow structure). Therefore, our method becomes valid when the DC width and elongate degree are calculated for each condition. In this study, we adopted a simplified accretion-flow structure; however, a study with a more realistic structure calculated by GRMHD simulations is left for future work \citep[e.g., ][]{Narayan2012, Ressler2015, Porth2019}. In addition, the proton and the electron temperatures in the optically thin accretion disk are thought to be different \citep{Narayan1995, Yuan2014}. Thus, it is also necessary to use the results of two-temperature GRMHD simulations \citep{Ressler2015, Mizuno2021}.


Our method 
requires observations with 
a sufficiently high-resolution 
to resolve the DC and the observation cadence of $\simeq 2\ t_{\rm g}$ (1 day for M87). The space-VLBI observation, such as "Black Hole Explorer (BHEX)" mission \citep{EHE_whitepaper, BHEX_SPIE1, BHEX_SPIE_Japan, BHEX_SPIE_Kawashima} is a promising candidate for such high-resolution observations. As for the observing duration, $10-20 \ t_{\rm g}$ (5-7 days for M87) is required in order to capture the full evolution of the flare, including its brightening and fading phases, as well as the accompanying shift of the intensity-weighted center caused by the delayed brightening of the photon ring.
More precisely, it is necessary to downsample the simulated images to match actual observations and also consider a realistic array configuration, but this will be addressed in future work.

\ack
We thank the anonymous referee for their fruitful comments and helpful suggestions to improve the manuscript. Our numerical simulations were conducted with the FUJITSU Supercomputer PRIMEHPC FX1000 and FUJITSU Server PRIMERGY GX2570 (Wisteria/BDEC-01) at the Information Technology Center, The University of Tokyo, and Cray XC50 and HPE Cray XD2000 at the Center for Computational Astrophysics, National Astronomical Observatory of Japan. This work was supported by Takahashi Industrial and Economic Research Foundation (MMT), JSPS KAKENHI Grant numbers 22J10256, 22KJ0382 (MMT), 23K03448 (TK), 21H04488 (KO), 25K01045 (MMT \& KO) and Multidisciplinary Cooperative Research Program in CCS, University of Tsukuba. This work was also supported (in part) by MEXT as ”Program for Promoting Researches on the Supercomputer Fugaku” (Toward a unified view of the universe: from large scale structures to planets; JPMXP1020200109, Structure and Evolution of the Universe Unraveled by Fusion of Simulation and AI; JPMXP1020230406) (KO), and by Joint Institute for Computational Fundamental Science (JICFuS, KO), Multidisciplinary Cooperative Research Program in CCS, University of Tsukuba.






\input{output.bbl}

\end{CJK}
\end{document}